# Surveying the ice condensation period at southern polar Mars using a CNN

**Gergácz Mira** [1,2,3,4]  **Kereszturi Ákos** [3,5]

[1] *ELTE Institute of Physics, H-1117 Budapest, Pázmány Péter 1/A, Hungary*

[2] *Wigner RCP, H-1121 Budapest, Konkoly-Thege Miklós 29-33, Hungary*

[3] *Konkoly Thege Miklos Astronomical Institute, HUN-REN Research Centre for Astronomy and Earth Sciences, H-1121 Budapest, Konkoly-Thege Miklós 15-16, Hungary*

[4] *Luleå University of Technology, Bengt Hultqvists väg 1, 981 92 Kiruna, Sweden*

[5] *CSFK, MTA Centre of Excellence, H-1121 Budapest, Konkoly-Thege Miklós 15-17, Hungary*

## Abstract

Before the seasonal polar ice cap starts to expand towards lower latitudes on Mars, small frost patches may condensate out during the cold night and they may remain on the surface even during the day in shady areas. If ice in these areas can persist before the arrival of the contiguous ice cap, they may remain after the recession of it too, until the irradiation increases and the ice is met with direct sunlight. In case these small patches form periodically at the same location, slow chemical changes might occur as well.

To see the spatial and temporal occurrence of such ice patches, large number of optical images should be searched for and checked. The aim of this study is to survey the ice condensation period on the surface with an automatized method using a Convolutional Neural Network (CNN) applied to High-Resolution Imaging Science Experiment (HiRISE) imagery from the Mars Reconnaissance Orbiter (MRO) mission. The CNN trained to recognise small (ranging from 0.5 to 300 m in diameter) ice patches is automatizing the search, making it feasible to analyse large datasets.

Previously a manual image analysis was conducted on 110 images from the southern hemisphere, captured by the HiRISE camera on-board the MRO space mission [1, 2]. Out of these, 37 images were identified with smaller ice patches, which were used to train the CNN. This approach is applied now to find further images with potential water ice patches in the latitude band between -40° and -60°, but contrarily to the training dataset recorded between 140-200° solar longitude, the images were taken from the condensation period between Ls = 0° to 90°. The model was ran on 171 new HiRISE images randomly picked from the given period between -40° and -60° latitude band, creating 73155 small image chunks. The model classified

2 images that show small, probably recently condensed frost patches and 327 chunks were predicted to show ice with more than 60% probability.

# 1 Introduction

In this study, a large amount of HiRISE images are surveyed using automatized methods to search for small scale ice patches on the surface.

Before the expansion of the seasonal polar ice cap, frost may condensate out on the surface in higher latitudes [3]. This happens overnight and the ice disappears as the Sun comes up and the temperature rises [4], however there is a possibility of some of these patches remaining on the surface in shaded areas. In the southern summer as the irradiance reaches these shaded areas too, the temperature may rise quickly enough for the liquid phase to emerge for a short period of time [5, 6, 7]. In this work such ice patches are looked for, using automatized survey of high resolution images taken by the High Resolution Imaging Science Experiment (HiRISE) camera on board of the Mars Reconnaissance Orbiter (MRO).

As the surface and atmosphere of Mars exhibits low inertia and thermal conductivity [8, 9], it is possible for small frost patches [10] to condensate out on the surface in locations where direct irradiation from the Sun doesn't reach them. These areas can be in the shadow of poleward-facing slopes, rocky fields or bottom of craters. Over time the frost in these places may also be exposed to direct sunlight which may cause a rapid temperature rise in the ice – it is not yet known whether a liquid phase [10, 11] may then appear, which is an important question for chemical transformations and the potential for life [12, 13, 14].

If the liquid phase emerges in these places recurrently, it might influence slow, low temperature chemical changes on Mars, especially if supported by subzero temperature microscopic liquid water like proposed for hydrogen-peroxide decomposition [15] or for sulphate formation [16]. Such locations might need focused analysis in the future by orbiters monitoring them, which requires specific information on their location, the time period in which ice is present there, and a selection of the best candidates among them regarding potential chemical changes.

Surveying HiRISE images with small frost patches on them requires an extensive amount of manual work, as ice should be separated from bright rocks, bright patches on slopes with solar facing exposure direction, and large part of the surface might be bright dust covered.. Additionally, other distinct surface features, occasionally captured under less than ideal conditions or affected by hazes or clouds, further complicate the selection process.

## 2  Methods

In this work an automated CNN model is used on 171 HiRISE images out of the available 1566 in the area of interest, to look for images with newly condensed ice patches visible.

### 2.1  The HiRISE camera

The MRO spacecraft has orbited Mars since 2006 with the HiRISE camera on-board, mapping about 4% of the surface by 2021, often revisiting specific areas. This high-resolution camera utilizes a 0.5-meter diameter mirror telescope, the largest ever used around another planet. Positioned 300 kilometres above Mars, it captures detailed images with a pixel size of 25 centimetres, enabling comprehensive surface surveys. The images cover 6 kilometres in width (20,000 pixels) and can be extended up to 60 kilometres in length (200,000 pixels). Operating between 14-16 local time, it produces colour images in three wavelength bands: 400-600 nm (blue-green, B-G), 550-850 nm (red, R), and 800-1000 nm (near-infrared, NIR) in the central field of view.

### 2.2  Surveyed region and dataset

The area of interest was the latitude band between -40° and -60° in the seasonal range of 0°-90° solar longitude, when water starts to condensate out on the surface in the southern hemisphere. A bright patch got identified as frost if it is located on the poleward side of a shading surface form, does not cast shadow and its edges are slightly diffuse.

### 2.3  Convolutional Neural Network (CNN)

Given the large size of the dataset and image dimensions, conducting manual analysis on all the available images is not feasible. Therefore a previously trained model is introduced, which was successful at recognising small ice patches left behind during the recession of the southern polar cap [1].

CNNs excel in classifying images and handling large datasets efficiently. They employ convolutional layers to capture local patterns and features by using convolutional kernels that scan the input image, extracting textures, edges, and gradients. These layers generate a feature map, showcasing kernel activations across spatial locations. Nonlinear activation functions enhance the network's ability to recognize intricate connections between these features, enabling nonlinear learning, which enables the network to recognise more complex features. Post-convolution, pooling operations reduce feature map dimensions, condensing information in local areas. This improves computational efficiency, allowing the network to prioritize feature presence over precise positional details within the input image.

All these above make CNNs an optimal tool for recognising the small features on the surface we are looking for in this paper. A HiRISE image was classified as 'good' if the average probability for an ice patch was over 60% across more than half of its sections, each with at least a 60% prediction for a small icy patch. If the icy sections were fewer than half, the image was still categorized as 'good' with an average probability above 20%. Images were labelled as 'hard to identify' when over half of the sections predicted ice with an average probability between 40-60%. If fewer than half but more than one section was considered icy, the minimum percentage for this category was set at 1%. Anything falling below this minimum threshold was classified as 'bad'.

## 3 The neural network

A smaller Xception network [17] was used in this work, trained on 18646 image chunks from previously categorised HiRISE images. Xception networks use depth wise separable convolution, instead of using the layers sequentially. This way the convolution operation is separated into two main steps, reducing the operations required and increasing the computational efficiency. The applied model can be seen in Figure 1.

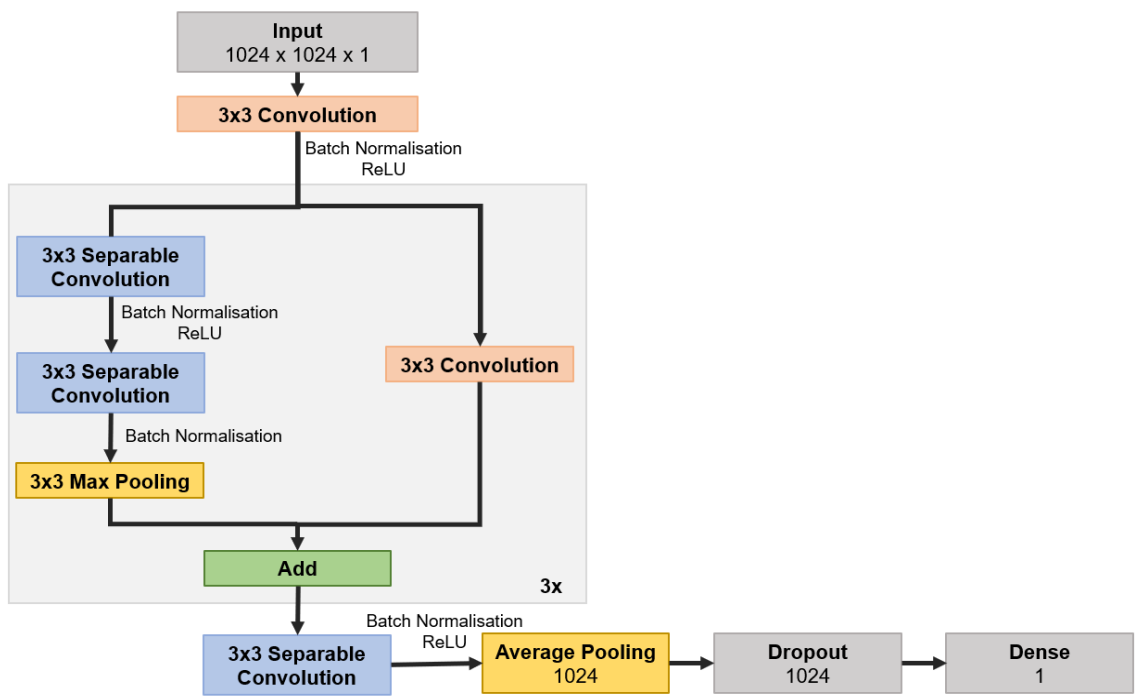

**Figure 1.** The structure of the model. Every ReLU activation is preceded by a Batch Normalisation, to normalise the inputs to the activation and improve learning [1].

The model was running on a batch size of 25 through 73 training epochs. During each epoch, the network processes the entire training dataset and updates its parameters to improve predictions. However, excessive epochs should be avoided to prevent overfitting, where the network learns noise and struggles to generalize with new data. To address this, a 3x3 sized kernel is utilized with a stride of 2, while a dropout rate of 0.5 is set to prevent the model from relying too heavily on specific nodes during prediction.

The network's learning process is supervised, utilizing labelled data during training to discern patterns distinguishing between different labels (such as 'good' or 'bad'). Validation involved 20% of the dataset, randomly sampled to ensure reliability. Given the network's task of classifying two image types, a sigmoid activation function was selected for its suitability in binary classification problems. Correspondingly, the loss function was set to binary crossentropy, commonly paired with sigmoid activations. Optimization employed the Adam optimizer [18]. This implementation was created using Keras with a Tensorflow backend [19, 20] in Python, and both training and testing were executed on a processing unit provided by the Wigner Scientific Computing Laboratory.

## 4 Training dataset

The model used in this work was trained to identify between two types of images: ones with small frost patches visible and ones with none. To achieve this, a set of previously categorised HiRISE images were used to teach the program distinguish between these two categories.

Due to the size of HiRISE images, they required segmentation into smaller portions before being used as input. This segmentation process was executed using the Mars Orbital Data Explorer Access software [21], responsible for downloading from the Mars Orbital Data Explorer (ODE) site [22], converting to greyscale, segmenting the images into chunks, and excluding chunks with black pixels surrounding the image centre, thereby disregarding black borders and most damaged images. Each chunk measures 1024 pixels by 1024 pixels, a size crucial for considering surface features during the training process, beyond just the bright patches themselves.

Following the chunking process, the initial training dataset of 72 images expanded significantly to encompass 18646 image chunks. Among these 20% exhibited noticeable small frost patches, whereas the remaining chunks displayed either no ice patches or featured $CO_2$ sublimation visible on the surface, as depicted in Figure 2 (specifically the 1st column, 3rd image). Figure 2 is a visualization representing a small section (1024 x 1024 pixels, equivalent to approximately 307 x 307 meters) of the dataset employed for training purposes.

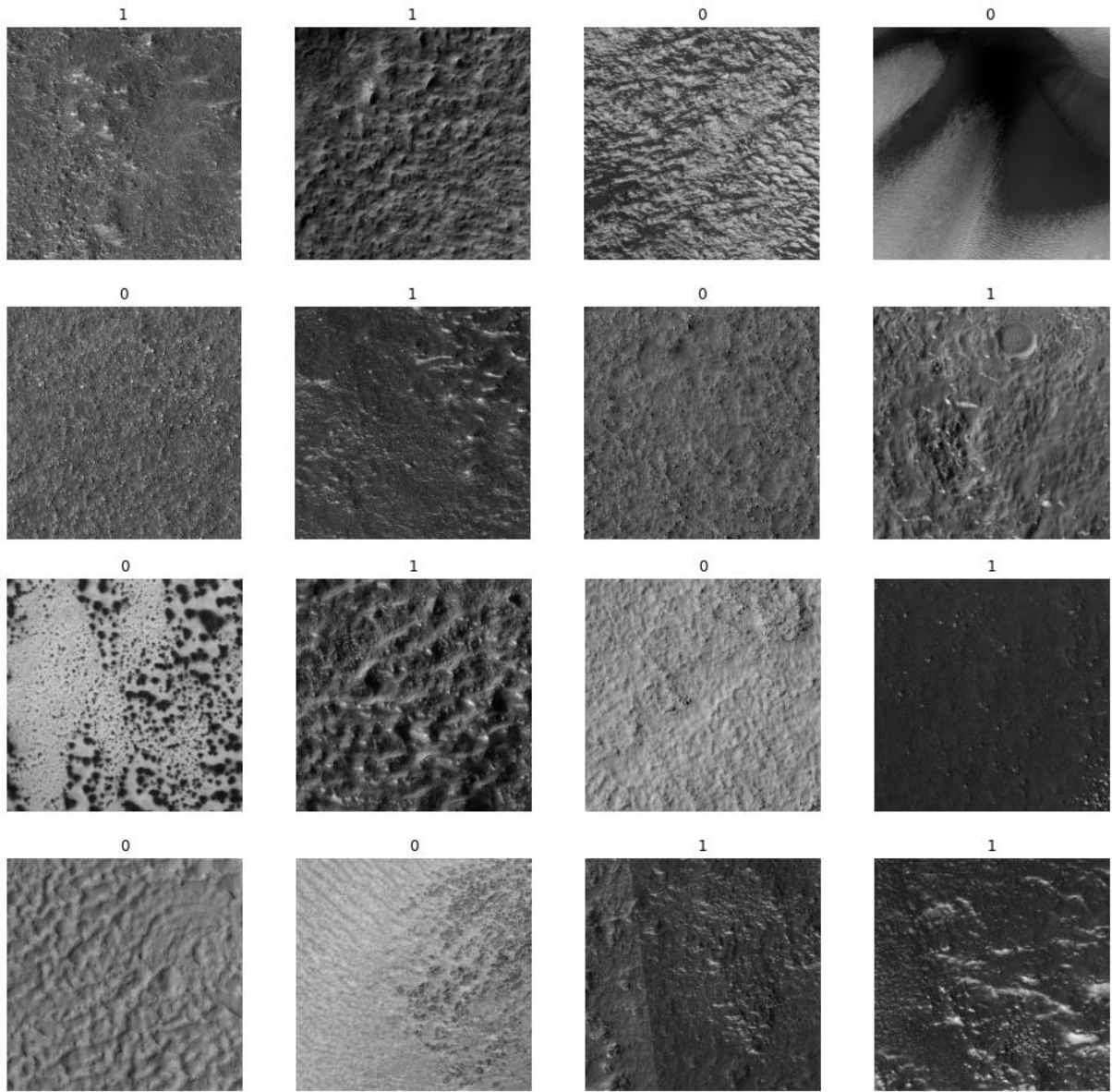

**Figure 2:** A few examples from the training dataset (1024 x 1024 pixel, e.g. 307 x 307 m sized) [1]. Images marked with 1 have small ice patches, the ones with 0 have no small frost patch or $CO_2$ ice is sublimating from an extended area of the surface causing dark and not bright patches.

## 5 Results

The trained neural network analysed 171 new HiRISE images (cut into 73155 chunks) from the area and seasonal range of interest. Small insets of 307 x 307 m sized areas are visualised in Figure 3 with their corresponding predictions for small ice patches being present. The program found faint frost patches with high probability, and effectively filtered out surfaces with no shading forms where ice could remain. Percentages between 40-60% were considered as 'hard to identify', like the one in image 8 on Figure 3.

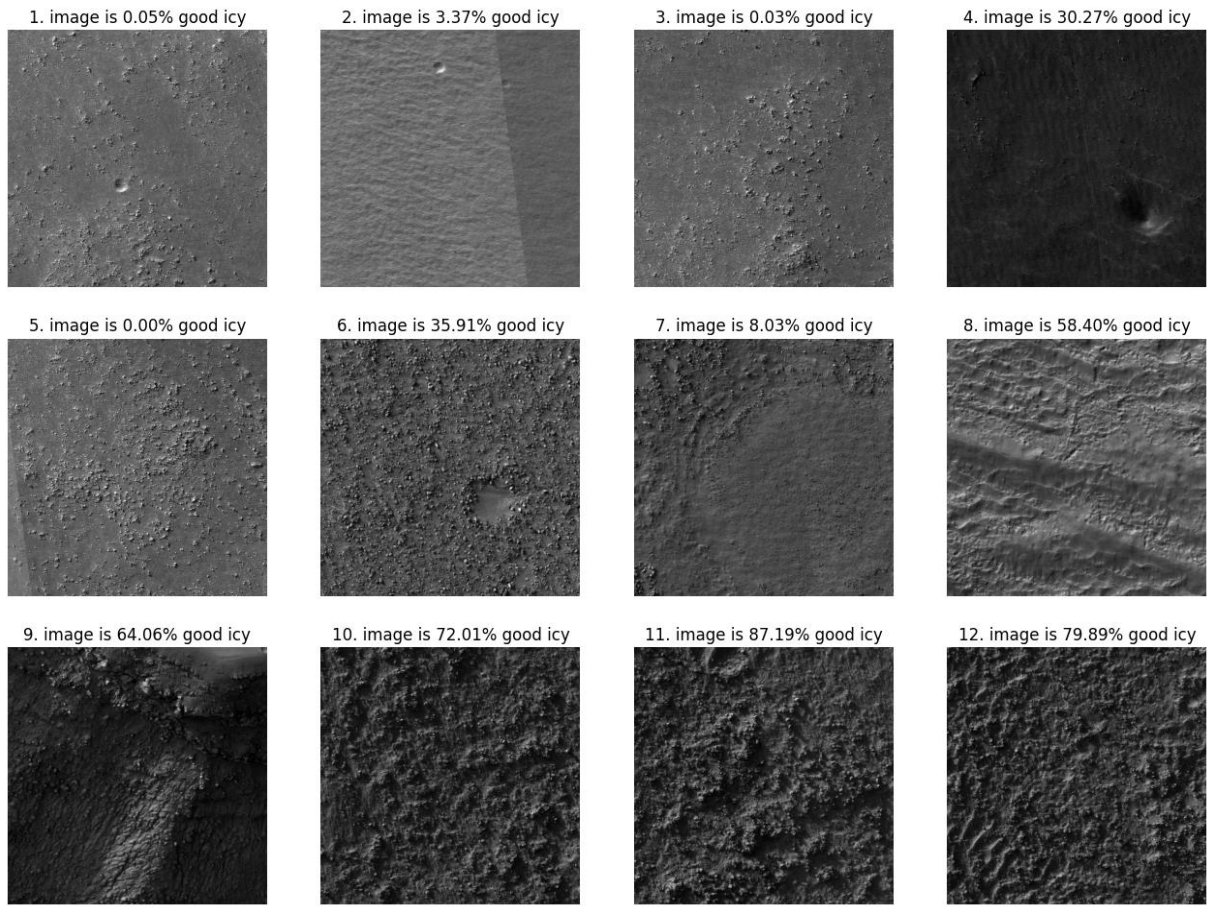

**Figure 3:** Example image chunks and their corresponding predicted probability for having small icy patches visible on them. Above each image the predicted probability of having small ice patches on the surface is displayed in a percentage. The model recognised faint frost patches on images 10-12, which were confirmed after colour analysis, however made a mistake on the 9. image, where no ice patches were visible even on the RGB image.

Out of the available 1566 images that were taken in the area and time of interest, 171 got analysed. These got cut into 73155 chunks, from which around 50 got manually checked and only 4 of predictions were incorrect. The common mistake on these chunks was the false classification of small, bright patches between rocks on the sunlit side. Out of the 171 images 2 got classified as 'good' by the program, 1 of them was not confirmed after manual analysis, when getting the colour information. 17 images got 'hard to identify' label, from which all were considered 'bad' after manual analysis. Image ESP_017092_1310 is the one with identified small ice patches on it, frost patches present around small (0.5 metre) diameter rocks, visible on the shaded side. Locations like this are common place for ice patches to remain on the surface after the recession of the seasonal polar ice cap too [2].

In Figure 3, 10-12. images have faint frost patches on the shaded side of the rocks, colour information was needed to verify these. On the image 9 bright patches between rocks, but on the sunlit side got misidentified as ice by the model. This was a common mistake made by the model on other chunks as well.

# 6  Discussion

The HiRISE image with identified frost patches was observed at Ls = 66° in -48° latitude, which time and area is before the spread of the continuous seasonal southern polar ice cap. It probably formed as non-continuous ice patch during night without expanding as a large areal frost cover. The image with sufficient frost patches present exhibited a rocky field consisting of small (0.5 metre) diameter rocks, which is where ice patches commonly remain after the recession of the southern polar ice cap [2].

The model seemed sufficient in filtering out HiRISE images with no shading surface forms or ice patches present, making it a useful tool to analyse more of the surface.

## Acknowledgement

The authors thank the Wigner Scientific Computing Laboratory for their support.